\documentstyle[nato]{crckapb} 
\input epsf

\begin{opening}
\title{Domain Wall Solutions}
\author{Tanmay Vachaspati}
\institute{Physics Department, Case Western Reserve University\\
           Cleveland, OH 44106, USA.}
\end{opening}

\runningtitle{Domain Walls}

\begin{document}

\section{Introduction}
\label{introduction}

Of the large variety of topological defects that are
known, domain walls are the simplest to understand in
many ways. The underlying topology that gives rise to
domain walls is discrete and can easily be visualized.
In some models the equations of motion can be solved 
analytically leading to explicit expressions. Since the
topology is discrete, formation via the Kibble mechanism
can be discussed in terms of known results in percolation
theory. Hence it is instructive to embark on a study of
topological defects starting with the simplest case of 
domain walls. The purpose of these lectures is to get
you started.

As often happens, the simplest of considerations evolve
into more complex phenomena. And domain walls are no
exception. Only in the simplest of models are domain
walls simple. As we go to somewhat more elaborate and
realistic models, the domain walls in them get more
complicated and realistic! Some of the developments
that I will describe have direct connection with the
investigation of domain walls in He-3 \cite{SalVol88}. In 
these lectures I will develop the subject enough so that 
you can see the complexities emerging. However, there
will not be enough time to cover all the recent advances
and those woods will be left for you to explore on your
own. An important part of the forest that I will not get
into is the cosmology of domain walls. This includes discussion
of their dynamics and cosmological evolution. There simply 
isn't enough time for it. I am hoping that the lectures
by Professor Arodz will rectify some of these omissions.

I will begin by describing the simplest of domain walls.
This is the ``kink'' in a $Z_2$ model. 
Then I will consider the more complicated field theoretic 
model based on $SU(5)\times Z_2$ symmetry. (The discussion is 
easily carried over to the case of $SU(N)\times Z_2$ for odd $N$.) 
The discrete symmetry responsible for domain walls is still 
$Z_2$ like in the simple model. Here, however, we will find 
a discrete spectrum of domain walls all having the same 
topology but different masses and other properties. The 
formation of domain walls in this model via the Kibble 
mechanism and the importance of these considerations in 
connection with cosmology will be discussed. 

\section{The kink}
\label{thekink}

Consider the $\phi^4$ Lagrangian in 1+1 dimensions labeled by $(t,z)$
\begin{equation}
L = {1\over 2}(\partial _\mu \phi )^2 - {\lambda\over 4} (\phi^2 -\eta^2 )^2
\label{z2model}
\end{equation}
where $\phi (t,z)$ is a real scalar field -- also called the
order parameter. The Lagrangian
is invariant under $\phi \rightarrow -\phi$ and
hence possesses a $Z_2$ symmetry. For this
reason, the potential has two minima: $\phi = \pm \eta$,
and the ``vacuum manifold'' has two-fold degeneracy.

Consider the possibility that $\phi = +\eta$ at
$z= +\infty$ and $\phi = -\eta$ at $z=-\infty$. In
this case, the continuous function $\phi (z)$ has
to go from $-\eta$ to $+\eta$ as $z$ is taken from
$-\infty$ to $+\infty$ and so must necessarily
pass through $\phi =0$. But then there is energy in
this field configuration since the
potential is non-zero when $\phi =0$. Also, this configuration
cannot relax to either of the two vacuum configurations, say
$\phi (z) = + \eta$, since that involves changing the
field over an infinite volume from $-\eta$ to $+\eta$,
which would cost an infinite amount of energy.

Another way to see this is to notice the presence of a
conserved current:
$$
\eta j^\mu = \epsilon^{\mu \nu} \partial_\nu \phi
$$
where $\mu, \nu =0,1$ and $\epsilon^{\mu \nu}$ is the
antisymmetric symbol in 2 dimensions. Clearly $j^\mu$
is conserved and so we have a conserved charge in the
model:
$$
\eta Q = \int dz j^0 = \phi (+\infty ) - \phi (-\infty ) \ .
$$
For the vacuum $Q=0$ and for the configuration described
above $Q=1$. So the configuration cannot relax into the
vacuum - it is in a different topological sector.

To get the field configuration with the boundary
conditions $\phi (\pm \infty ) =\pm \eta$, one would have
to solve the field equation resulting from the Lagrangian
(\ref{z2model}). This would be a second order differential equation.
Instead, one can use the clever method first derived by
Bogomolnyi \cite{Bog76} and obtain a first order differential
equation. The method uses the energy functional:
\begin{eqnarray*}
E & = & \int dz
     \left [ {1\over 2}(\partial_t \phi)^2 + 
           {1\over 2}(\partial_z \phi)^2 + V(\phi ) \right ] \\
& = &  \int dz  \left [ {1\over 2}(\partial_t \phi)^2 +
                {1\over 2}(\partial_z \phi - \sqrt{2V(\phi )} ~ )^2 +
                  \sqrt{2V(\phi )}\partial_z \phi \right ]  \\
& = & \int dz \left [ {1\over 2}(\partial_t \phi)^2 +
{1\over 2} (\partial_z \phi - \sqrt{2V(\phi )} ~ )^2 \right ] +
\int^{\phi(+\infty )}_{\phi(-\infty )} d\phi ' \sqrt{2V(\phi ' )} \\
\end{eqnarray*}
Then, for fixed values of $\phi$ at $\pm \infty$, the energy is
minimized if
$$
\partial_t \phi =0
$$
and
$$
\partial_z \phi - \sqrt{2V(\phi )} = 0 \ .
$$
Furthermore, the minimum value of the energy is:
$$
E_{min} =  \int^{\phi(+\infty )}_{\phi(-\infty )}
d\phi ' \sqrt{2V(\phi ' )} \ .
$$
In our case,
$$
\sqrt{V(\phi )} = \sqrt{{\lambda\over 4}} (\eta^2 -\phi^2  )
$$
which can be inserted in the above equations to get the ``kink''
solution:
$$
\phi = \eta ~ {\rm tanh} \biggl (\sqrt{\lambda\over 2} \eta z \biggr )
$$
for which the energy per unit area is:
\begin{equation}
\sigma_{kink} = {{2\sqrt{2}} \over 3} \sqrt{\lambda} \eta^3
              = {1 \over 3} {{m^3}\over {\sqrt{\lambda}}}
\label{sigmakink}
\end{equation}
where $m = \sqrt{2\lambda} \eta$ is the mass of excitations 
(particles) of $\phi$ in the vacuum of the model.
Note that the energy density is localized in the region where
$\phi$ is not in the vacuum, {\it i.e.} in a region of thickness
$\sim m^{-1}$ around $z=0$.

We can extend the model in eq. (\ref{z2model}) to 3+1 dimensions and
consider the case when $\phi$ only depends on $z$ but not on $x$ and
$y$. We can still obtain the kink solution for every value of $x$ and
$y$ and so the kink solution will describe a ``domain wall'' in the
$xy-$plane.

At the center of the kink, $\phi =0$, and hence the $Z_2$
symmetry is restored in the core of the kink. In this
sense, the kink is a ``relic'' of the symmetric phase
of the system. If kinks were present in the universe
today, their interiors would give us a glimpse of what
the universe was like prior to the phase transition.

\section{$SU(5)\times Z_2$ walls}
\label{suNZ2walls}

An example that is more relevant to cosmology is motivated
by Grand Unification. Here we will consider the $SU(5)$ model:
\begin{equation}
L = {\rm Tr} ( D_\mu \Phi )^2 - 
      {1\over 2} {\rm Tr} (X_{\mu\nu} X^{\mu\nu}) -V(\Phi )
\label{su5bosoniclagrangian}
\end{equation}
where, in terms of components, $\Phi = \Phi^a T^a$ is an
$SU(5)$ adjoint, the gauge field strengths are 
$X_{\mu\nu} = X_{\mu\nu}^a T^a$ and the $SU(5)$ generators
$T^a$ are normalized such that ${\rm Tr} (T^aT^b) = \delta^{ab}/2$.
The definition of the covariant derivative is:
\begin{equation}
D_\mu \Phi = \partial_\mu \Phi -ie [X_\mu , \Phi]
\label{covariantderiv}
\end{equation}
and the potential is the most general quartic in $\Phi$:
\begin{equation}
V(\Phi ) = -m^2 {\rm Tr} (\Phi^2) + h [{\rm Tr} (\Phi^2)]^2 
   + \lambda {\rm Tr} (\Phi^4) + \gamma {\rm Tr} (\Phi^3) - V_0\ ,
\label{su5potential}
\end{equation}
where, $V_0$ is a constant that we will choose so as to set
the minimum value of the potential to zero.

The $SU(5)$ symmetry is broken to 
\begin{equation}
H = [SU(3)\times SU(2)\times U(1)]/Z_6
\label{unbrokenH}
\end{equation}
if the Higgs acquires a VEV equal to
\begin{equation}
\Phi_0 = {\eta\over {2\sqrt{15}}} {\rm diag} (2,2,2,-3,-3)
\label{phi0}
\end{equation}
where
\begin{equation}
\eta = {m \over {\sqrt{\lambda '}}} \ , \ \ \ 
\lambda ' \equiv h+ {7\over {30}}\lambda \ .
\label{lambdaprime}
\end{equation}
For the potential to have its global minimum at $\Phi =\Phi_0$,
the parameters are constrained to satisfy:
\begin{equation}
\lambda \ge 0 \ , \ \ \ \lambda ' \ge 0 \ .
\label{constraint1}
\end{equation}
For the global minimum to have $V(\Phi_0 )=0$, in
eq. (\ref{su5potential}) we set
\begin{equation}
V_0 = - {{\lambda '}\over 4} \eta^4 \ .
\label{V0}
\end{equation}

The model in eq. (\ref{su5bosoniclagrangian}) with the potential 
in eq. (\ref{su5potential}) does not have any topological
domain walls because there are no broken discrete symmetries. In
particular, the $Z_2$ symmetry under $\Phi \rightarrow -\Phi$ 
is absent due to the cubic term. However if $\gamma$ is small, 
there are walls connecting the two vacua related by 
$\Phi \rightarrow -\Phi$ that are almost topological. 
In our analysis we will set $\gamma =0$, in which case the symmetry 
of the model is $SU(5)\times Z_2$. An expectation of $\Phi$ breaks 
the $Z_2$ symmetry and leads to topological domain walls.

The Lagrangian in eq. (\ref{su5bosoniclagrangian}) provides us
with the (second order) equations of motion for all the fields. 
However, it does not tell us the boundary conditions on the
fields. The boundary conditions will depend on the class of 
solutions that we are interested in. This is the first issue
that we need to settle.

We would like a solution that has the correct topology.
Topology tells us that the vacuum manifold is disconnected.
One of the disconnected regions is described by 
$- U^\dag \Phi_0 U$ and the other by $+ U^\dag \Phi_0 U$,
where $U \in SU(5)$. To obtain solutions we need 
$\Phi_- \equiv \Phi (-\infty )$ to be in the first sector and
$\Phi_+ \equiv \Phi (+\infty )$ to be in the second disconnected
sector. Since we can globally rotate the fields by any $SU(5)$ 
transformation, we are free to set $\Phi_- = -\Phi_0$. But we 
still have to choose $\Phi_+ =  +U^\dag \Phi_0 U$ for some
$U \in SU(5)$.

The freedom to choose $\Phi_+$ is greatly reduced
by the following result: a static solution to the field 
equations is only possible if $[\Phi_- , \Phi_+ ] = 0$.
I will not give the proof here but you can find it in
the Appendix of Ref. \cite{Vac01}.

The condition $[\Phi_- , \Phi_+ ] = 0$ still permits
a large variety of boundary conditions $\Phi_+$. But
the different choices for $\Phi_+$ can all be rotated
into a diagonal choice by a transformation in the 
unbroken global symmetry group which leaves $\Phi_-$ 
invariant. Hence we need only consider a discrete
set of choices for $\Phi_+$, and these are:
\begin{equation}
\Phi_+^{(0)}=
\epsilon {\eta\over {2\sqrt{15}}} {\rm diag} (2,2,2,-3,-3)
\label{Phi+0}
\end{equation}
\begin{equation}
\Phi_+^{(1)}=
\epsilon {\eta\over {2\sqrt{15}}} {\rm diag} (2,2,-3,2,-3)
\label{Phi+1}
\end{equation}
\begin{equation}
\Phi_+^{(2)}=
\epsilon {\eta\over {2\sqrt{15}}} {\rm diag} (2,-3,-3,2,2)
\label{Phi+2}
\end{equation}
The superscript on $\Phi_+$ corresponds to the number
of entries of $\Phi_-$ that have been permuted. The
parameter $\epsilon$ has the value $-1$ for the topological
domain walls. Later we will find that non-trivial solutions
also exist for $\epsilon =+1$. These correspond to non-topological
domain walls.

It is tempting to try to obtain the domain wall solutions
using Bogomolnyi's method. This has not been done and remains 
an interesting open problem. However, we can write down the 
solution for the zero permutation case since it is exactly like 
the $Z_2$ case:
\begin{equation}
\Phi^{(0)} = \tanh (X) \Phi_0
\label{su5kink}
\end{equation}
where $X \equiv mx/\sqrt{2}$.

The other cases are a little more involved but still
quite simple. The solution is of the form:
\begin{equation}
\Phi_k = F_+ (x) {\bf M_+} + F_- (x) {\bf M_-} + g(x) {\bf M}\ ,
\label{kinkexplicit2}
\end{equation}
where
\begin{equation}
{\bf M}_+ =  {{\Phi_+ + \Phi_-}\over {2}} \ , \ \ {\bf M}_- =
{{\Phi_+ - \Phi_-}\over {2}} \label{M+M-} \ ,
\end{equation}
$g(\pm \infty)=0$ and ${\bf M}$ is yet to be found.

It is now convenient to move from the $SU(5)$ to the 
$SU(N)$ case. Let $N=2n+1$, so $n=2$ when $N=5$. Also
let us label the number of permutations in the boundary
conditions by the integer $q$. For $N=5$, the three
possible boundary conditions in eq. (\ref{Phi+0})-(\ref{Phi+2})
correspond to $q=0,1,2$. The advantage of dealing with
the general case is that we can write one equation in
terms of $n$ and $q$ rather than 3 equations, one per case.

The formulae for ${\bf M}_\pm$ can now be explicitly
written (we set $\epsilon =-1$ for now): 
\begin{equation}
{\bf M}_+ = \eta N \sqrt{1 \over {2N(N^2-1)}} 
 {\rm diag} ( {\bf 0}_{n+1-q}, {\bf 1}_q, 
                 -{\bf 1}_q, {\bf 0}_{n-q} )
\label{M+} \ ,
\end{equation}
\begin{equation}
{\bf M}_- = \eta  \sqrt{1 \over {2N(N^2-1)}} 
{\rm diag} ( -2n {\bf 1}_{n+1-q}, {\bf 1}_q,
                {\bf 1}_q, 2(n+1){\bf 1}_{n-q} ) \ .
\label{M-}
\end{equation}
We have used ${\bf 0}_k$ and ${\bf 1}_k$ to denote
the $k-$dimensional zero and unit matrices respectively.
Note that the matrices ${\bf M}_\pm$ are orthogonal:
\begin{equation}
{\rm Tr}({\bf M}_+{\bf M}_-) = 0 \ , \label{trM+M-}
\end{equation}
but are not normalized to 1/2. The boundary conditions for $F_\pm$
are:
\begin{eqnarray}
F_- (- \infty ) &=& -1 \ , \ \  F_- (+\infty ) =+1 \ , \nonumber \\
F_+ (- \infty ) &=& +1 \ , \ \  F_+ (+\infty ) =+1 \ .
\label{Fpmbc}
\end{eqnarray}
The advantage of this form of the ansatz is that, for particular
values of the parameters of a quartic potential in the $q=n$
topological ($\epsilon =-1$) case, one finds the explicit and
simple solution $F_+ (x) =1$, $F_- (x) = \tanh (\sigma x)$ and
$g(x)=0$, where $\sigma =mx/\sqrt{2}$ \cite{PogVac00,Vac01}. 
Also, for $q=0$, $\epsilon =-1$, the solution is the embedded 
$Z_2$ kink {\it i.e.} $F_+(x)=g(x)=0$, $F_- (x) = \tanh (\sigma x)$.

Now we would like to find the unknown matrix ${\bf M}$ in the ansatz
(\ref{kinkexplicit2}). By inserting the ansatz in the equations
of motion \cite{PogVac01}, it turns out that ${\bf M}$ is uniquely 
determined and this is: 
\begin{eqnarray}
{\bf M} = \mu \, {\rm diag} &(& q(n-q){\bf 1}_{n+1-q}, 
   -(n-q)(n+1-q){\bf 1}_q , 
\nonumber \\ 
   &-&(n-q)(n+1-q){\bf 1}_q, q(n+1-q){\bf 1}_{n-q} )
\label{Mresult}
\end{eqnarray}
with $\mu$ being a normalization factor in which we also include
the energy scale $\eta$ for convenience:
\begin{equation}
\mu = \eta [ 2q(n-q)(n+1-q)\{ 2n(n+1-q)-q\} ]^{-1/2} \ .
\label{muvalue}
\end{equation}
Note that the matrix ${\bf M}$ is not normalizable if $q=0$ or if
$q=n$. For these values of $q$, we can set $g(x)=0$.

\begin{figure}
\center{\epsfxsize= 0.6\hsize\epsfbox{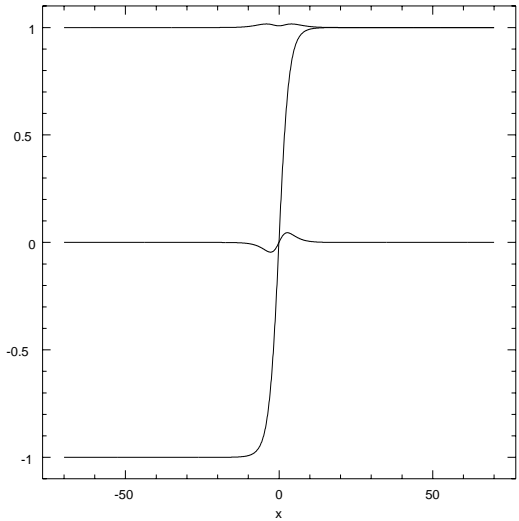}}
\caption{\label{qeq1top} The profile functions $F_+$ (nearly
1 throughout), $F_-$ (shaped like a $\tanh$ function), and $g$
(nearly zero) for the $q=1$ topological kink with parameters
$h=-3/70$, $\lambda=1$ and $\eta =1$.}
\end{figure}

The functions $F_\pm (x)$ and $g(x)$ can be found by solving their
equations of motion derived from the Lagrangian together with the
specified boundary conditions. There is no guarantee that a solution
will exist and so we find the solutions explicitly for $N=5$ with
a quartic potential. The solution for $q=0$ is simply
that in eq. (\ref{su5kink}). The $q=1$ profile functions are 
evaluated numerically and shown in Figure \ref{qeq1top}. The 
profiles for $q=2$ have also been evaluated numerically 
in Ref. \cite{PogVac00}. However, an interesting situation
arises for $q=2$ when 
\begin{equation}
{h\over \lambda} = -{{3}\over {N(N-1)}} = -{3\over {20}}
\label{critparameters}
\end{equation}
{}For this special value of coupling constants, the profile 
functions can be written as \cite{Vac01}:
\begin{equation}
F_+ =1 \ , \ \ F_- = \tanh (X) \ , \ \ g =0
\label{q=2analytic}
\end{equation}
It is not understood what is special about $h/\lambda =-3/20$
that it allows an analytic solution. Could it be that there
is a symmetry of the system at this point in parameter
space that simplifies matters? Could a Bogomolnyi type
analysis be done at this point?

The stability of these solutions has been investigated
in Ref. \cite{PogVac00,Vac01,PogVac01} and the $q=2$ solution
turns out to be the minimum energy domain wall solution. For
the particular choice of model parameters given in 
eq. (\ref{critparameters}) (valid for $N > 3$) \cite{Vac01}, 
the energy of the $q=2$ kink is:
\begin{equation}
\sigma = {{2\sqrt{2}}\over 3}{{m^3}\over \lambda}
            \left ( {{N-1}\over {N-3}} \right ) \ .
\label{sigma}
\end{equation}

An interesting point to note is that the ansatz is valid even if
$\Phi_\pm$ are not in distinct topological sectors {\it i.e.} even
if $\epsilon=+1$. These imply the existence of non-topological
kink solutions in the model. If we include a subscript $NT$ to
denote ``non-topological'' and $T$ to denote ``topological'', we
have
\begin{equation}
\Phi_{NTk} = F_+ (x) {\bf M}_{NT+} + F_- (x) {\bf M}_{NT-}
          + g(x) {\bf M}_{NT} \ .
\label{NTfinalansatz}
\end{equation}
Since $\Phi_{NT+}=-\Phi_{T+}$, we find
\begin{equation}
{\bf M}_{NT+} = {\bf M}_{T-} \ , \ \ {\bf M}_{NT-} = {\bf M}_{T+}
\ , \ \ {\bf M}_{NT} ={\bf M}_{T} \ . \label{NTtoT}
\end{equation}
Hence
\begin{equation}
\Phi_{NTk} = F_- (x) {\bf M}_{T+} + F_+ (x) {\bf M}_{T-}
          + g(x) {\bf M}_T \ .
\label{NTsolution}
\end{equation}
So to get $F_-$ ($F_+$) for the non-topological kink we have to
solve the topological $F_+$ ($F_-$) equation of motion with the
boundary conditions for $F_-$ ($F_+$). To obtain $g$ for the
non-topological kink, we need to interchange $F_+$ and $F_-$ in
the topological equation of motion. The boundary conditions for
$g$ are unchanged. Generally the non-topological solutions, if 
they exist, will be unstable. However, the possibility that some 
of them may be locally stable for certain potentials cannot be 
excluded.


\subsection{Domain wall lattice}

The forces between domain walls of different kinds has been 
studied by Pogosian \cite{Pog02}. What is most interesting
is that the $q=2$ wall and anti-wall can repel. Here one
needs to be careful about the meaning of an ``anti-wall''.
An anti-wall should have a topological charge that is
opposite to that of a wall. That is, a wall and its anti-wall
together should be in the trivial topological sector. But
this condition still leaves open several different kinds
of anti-walls for a given wall. To be specific consider
a domain wall across which the Higgs field changes from
$(2,2,2,-3,-3)$ to $-(2,-3,-3,2,2)$. (I am surpressing the
normalization factor for convenience of writing.) There
can be two corresponding anti-walls. In the first type
-- called Type I -- 
the Higgs field can go from $-(2,-3,-3,2,2)$ to
$+(2,2,2,-3,-3)$, thus reverting the change across
the first domain wall. In the second type (Type II), the Higgs 
field can go from $-(2,-3,-3,2,2)$ to $+(-3,2,2,-3,2)$.
Pogosian found that the force between a wall
and its Type I anti-wall is attractive, but the force
between a wall and its Type II anti-wall is repulsive.
This is understood by noting that the force between 
walls is proportional to ${\rm Tr} (Q_1Q_2)$ where
$Q_1$ and $Q_2$ are the charge matrices of the two
walls. If the Higgs field to the left side of the walls
is $\Phi_L$, between the two walls is $\Phi_0$, and is
$\Phi_R$ on the right-hand side of the walls, then
$Q_1 \propto \Phi_0 -\Phi_L$ and
$Q_2 \propto \Phi_R -\Phi_0$. Then the stable ($q=2$)
walls can have charge matrices proportional to
$(-4,1,1,1,1)$, $(1,-4,1,1,1)$, $(1,1,-4,1,1)$,
$(1,1,1,-4,1)$ and $(1,1,1,1,-4)$. (Stable anti-walls
have the same charges but with a minus sign.) Then
it is easy to see that one can take a wall with one
of the 5 charges listed above and it will repel
an anti-wall which has the $-4$ occurring in some
other entry. Hence there are combinations of
walls and anti-walls for which the interaction is
repulsive.

We know that topology forces a wall to be followed by
an anti-wall. Then we can set up a sequence of walls
and anti-walls in the following way:
\begin{equation}
...
Q^{(1)}{\bar Q}^{(5)}Q^{(3)}{\bar Q}^{(1)}Q^{(5)}{\bar Q}^{(3)}
...
\label{domainwallsequence1}
\end{equation}
where $Q_i$ and ${\bar Q}_i$ refer to a wall and an
anti-wall of type $i$ respectively. Alternately, this sequence 
of walls would be achieved with the following sequence of Higgs 
field expectation values:
\begin{eqnarray}
... \rightarrow -(2,2,2,-3,-3) 
              &\rightarrow& +(2,-3,-3,2,2) \nonumber \\
              &\rightarrow& -(-3,2,2,-3,2) \nonumber \\
              &\rightarrow& +(2,-3,2,2,-3) \nonumber \\
              &\rightarrow& -(2,2,-3,-3,2) \nonumber \\
              &\rightarrow& +(-3,-3,2,2,2) \nonumber \\
              &\rightarrow& -(2,2,2,-3,-3)
                                 \rightarrow ...
\label{Higgs sequence}
\end{eqnarray}
The forces between walls fall off exponentially
fast and hence the dominant forces will be between nearest 
neighbors. 

Note that the sequence of walls is periodic with a
period of 6 walls, and these 6 walls have a net topological
charge that vanishes. Hence we can put the sequence in
a periodic box i.e. compact space. This gives us a finite
lattice of domain walls.

The sequence described above is one whose period is 
the minimum possible (namely, 6). It is easy to construct
other sequences with greater periodicity. For example:
\begin{equation}
...
Q^{(1)}{\bar Q}^{(5)}Q^{(3)}{\bar Q}^{(4)}Q^{(2)}
{\bar Q}^{(1)}Q^{(5)}{\bar Q}^{(3)}Q^{(4)}{\bar Q}^{(2)}
...
\label{domainwallsequence2}
\end{equation}
is a repeating sequence of 10 domain walls. 

In ongoing work, we have checked that the lattice is a
solution of the equations of motion. However, it is not
stable. The instability is towards rotations of the walls
in internal space. For example, the ${\bar Q}^{(5)}$ antiwall
could rotate to become either a ${\bar Q}^{(1)}$ or a
${\bar Q}^{(3)}$ antiwall. Once this rotation takes place,
this antiwall will be attracted by the $Q^{(1)}$ or the
$Q^{(3)}$ neighbor, and annihilation can occur. An analogy
of this process is due to Pogosian -- if we place several
bar magnets along a line such that the North poles of
neighboring magnets face each other, then each bar magnet 
will be unstable to rotation. Once the bar magnets rotate and 
North poles face neighboring South poles, the magnets 
will attract and the chain of magnets is unstable to collapse. 
In this analogy, the rotation is in physical space; in the 
case of domain walls, the rotation is in internal space. 

The understanding of the domain wall lattice instability,
provides us with another avenue of investigation. If the
bar magnets were confined to one dimensions (for example,
they could be placed in a tube, or attached to a wire),
then the instability would cease to exist since rotations
would not be possible. The corresponding situation in the
case of domain walls is if the unbroken symmetry is too
restrictive to allow rotation. This can be achieved by
breaking the $SU(5)\times Z_2$ to $U(1)^4$ or something
smaller. A simpler alternative is to start out with a
model in which there are no rotational degrees of
freedom present. This scheme is discussed in \cite{PogVac02}.
One starts with a model that only has a discrete 
symmetry group that corresponds to permutations of the
diagonal entries of $SU(5)$ and the sign flip given
by $Z_2$. It can be shown that the domain wall lattice
is stable in this model.

\subsection{Formation of domain walls}
\label{formationofwalls}

The properties of the network of domain walls at formation has 
been determined by numerical simulations. The idea behind the
simulations is that the
vacuum in any correlated region of space is determined
at random. Then, if there are only two degenerate vacuua
(call them $+$ and $-$), there will be spatial regions 
that will be in the $+$ phase and others in the $-$ phase.
The boundaries between these regions of different phases is 
the location of the domain walls. This is nothing other than
the ``Kibble mechanism''.

\begin{table}[t]
\caption{
Size distribution of + clusters found by simulations on a
cubic lattice.
}\vspace{0.4cm}
\begin{center}
\begin{tabular*}{12.0cm}{|c@{\extracolsep{\fill}}ccccccc|}
\hline
{Cluster size} & {1} & {2} & {3} & {4} & {6} & {10} & {31082} \\[0.20cm]
\hline
{Number} & {462} & {84} & {14} & {13} & {1} & {1} & {1} \\[0.20cm]
\hline
\end{tabular*}
\end{center}
\end{table}

By performing numerical simulations, the 
statistics shown in Table I was obtained \cite{VacVil84}.
The data shows that there is essentially one giant connected $+$ cluster.
By symmetry there will be one connected $-$ cluster. In the infinite
volume limit, these clusters will also be infinite and their surface
areas will also be infinite. Therefore the topological domain wall
formed at the phase transition will be infinite.

What does the Kibble mechanism predict for $SU(5)\times Z_2$
domain walls? Once again we have to throw down values of the
Higgs field on a lattice and then examine the walls that would
form at the interface. We have found that there are 3 kinds of
wall solutions and so each one will be formed with some probability.
In the Kibble mechanism approach, the probability that a certain 
wall will form is directly related to the number of boundary values
that result in the formation of a defect. So we need to evaluate
all the boundary conditions that will lead to domain walls with
a certain value of $q$. In other words, we want to determine
the ``space of kinks'' for a fixed value of $q$ \cite{Vac01,PogVac01}.

Let the Higgs field to the left and right of a domain wall
be $\Phi_L$ and $\Phi_R$ respectively. Without loss of generalization,
we can take $\Phi_L = \Phi_0$. To start, let us find all values of
$\Phi_R$ that will give a $q=0$ kink. But there is only one such
value, namely $\Phi_R =-\Phi_0$. So the space of the $q=0$ kink
is just one point. Next, let us determine all $\Phi_R$ that will
give a $q=1$ kink. Here there are several possibilities since
different entries of $\Phi_0$ can be permuted to give different
$\Phi_R$. In fact, one can take any choice of $\Phi_R$, for
example along $-(2,2,-3,2,-3)$ and act by gauge transformations
belonging to the unbroken symmetry group $H$ 
(see eq. (\ref{unbrokenH})), and this will lead to another
choice for $\Phi_R$. Of course, some of these rotations will
leave $\Phi_R$ unchanged and these should not be counted. So
the space of $q=1$ kinks is given by the coset space:
\begin{equation}
\Sigma_1 = H/K_1 
\label{sigma1} 
\end{equation}
where
\begin{equation}
K_1 = [SU(2) \times U(1)^3]/Z_2  \ .
\label{k1forsu5} 
\end{equation}

A similar argument shows that the space of $q=2$ kinks is given
by:
\begin{equation}
\Sigma_2 = H/K_2 \ ,
\label{sigma2} 
\end{equation}
where
\begin{equation}
K_2 = [SU(2)^2 \times U(1)^2]/Z_2^2 \ .
\label{k2forsu5}
\end{equation}

Now that we have the space of kinks for each $q$, we note that
the space of the $q=0$ kink is zero dimensional, the space of
the $q=1$ kink is 6 dimensional, and the space of the
$q=2$ kink is 4 dimensional. Hence, in the Kibble mechanism
approach, the probability of a kink being of the $q=0$ or $q=2$ 
variety is zero, and the probability of the kink being of 
the $q=1$ variety is 1.

A subtlety that has not been discussed above is that there
is also the possibility that if we lay down Higgs fields
randomly, we may get $[\Phi_L,\Phi_R]\ne 0$. In this case,
as described earlier, there will be no static solution to the 
equations. Then the field configuration will evolve towards
a static configuration. Our discussion above assumes that
such a configuration has been reached, and neighboring domains
always have values of $\Phi$ that commute. This is not completely
satisfactory since there will be time scales that are associated
with the relaxation and these must be compared to other time
scales characterizing the phase transition.

To summarize, the Kibble mechanism predicts that only $q=1$ domain
walls will be formed at the $SU(5)\times Z_2$ phase transition.
However, we know that the stable variety of walls have $q=2$,
and the $q=1$ walls will decay into them. Precisely how the
$q=2$ walls convert into $q=1$ walls during a phase transformation
has not been studied. Nor is it known how long this relaxation
will take. The answers to these questions is of some importance
as we will briefly see in the next section.

\subsection{Importance in cosmology}
\label{importance}

The $SU(5)\times Z_2$ symmetry breaking also leads to topological
magnetic monopoles which are cosmologically unacceptable. However
these magnetic monopoles interact with the domain walls that we
have discussed in the earlier sections. If the domain wall is of
the $q=0$ variety, the monopole is certainly destroyed and its
magnetic charge spreads out on the domain wall
\cite{DvaLiuVac97,PogVac00}. If the domain wall
is of the $q=1$ variety, monopoles in a large number of orientations 
in the group space get destroyed whereas others survive. If the
domain wall is of the $q=2$ variety, a smaller number of monopoles
get destroyed \cite{PogVac01}. So the cosmological predictions
depend on the kinds of walls that are formed and their interactions
with magnetic monopoles. This is why it is important to consider
the types of domain walls that will be formed and the eventual
relaxation of the system, especially in regard to the number
density of magnetic monopoles.


\begin{thebibliography}{}
\bibitem[\protect\citeauthoryear{Salomaa and Volovik}{1988}]{SalVol88} 
M.M. Salomaa and  G.E. Volovik,
``Cosmiclike domain walls in superfluid $^3$He-B: 
Instantons and diabolical points in ($\vec k$, $  \vec r\, $) space'',
Phys. Rev. B {\bf 37}, 9298 - 9311 (1988).
\bibitem[\protect\citeauthoryear{Bogomolnyi}{1976}]{Bog76} 
E. B. Bogomolnyi, {\it Sov. J. Nucl. Phys.} {\bf 24}
449 (1976); reprinted in ``Solitons and Particles'', eds. C. Rebbi
and G. Soliani (World Scientific, Singapore, 1984).
\bibitem[\protect\citeauthoryear{Dvali, Liu and Vachaspati}{1997}]
{DvaLiuVac97}
G. Dvali, H. Liu and T. Vachaspati, 
{\it Phys. Rev. Lett.} {\bf 80}, 2281 (1998).
\bibitem[\protect\citeauthoryear{Vachaspati}{2001}]{Vac01}
T. Vachaspati, Phys. Rev. {\bf D63}, 105010 (2001).
\bibitem[\protect\citeauthoryear{Pogosian and Vachaspati}{2000}]
{PogVac00}
L. Pogosian and T. Vachaspati,
Phys. Rev. {\bf D62}, 123506 (2000). 
\bibitem[\protect\citeauthoryear{Pogosian and Vachaspati}{2001}]
{PogVac01}
L. Pogosian and T. Vachaspati,
Phys. Rev. {\bf D64}, 105023 (2001). 
\bibitem[\protect\citeauthoryear{Pogosian}{2002}]{Pog02}
L. Pogosian,
Phys. Rev. {\bf D65}, 065023 (2002).
\bibitem[\protect\citeauthoryear{Pogosian and Vachaspati}{2002}]{PogVac02}
L. Pogosian and T. Vachaspati, hep-th/0210232 (2002).
\bibitem[\protect\citeauthoryear{Vachaspati and Vilenkin}{1984}]
{VacVil84}
T. Vachaspati and A. Vilenkin,
Phys. Rev. {\bf {D30}}, 2036 (1984).

\end{thebibliography}
\end{document}